# Zero Permeability Materials ($Z\mu M$): A way out of the "Left Handed Materials" trap

N. Garcia, M. Muñoz and E. V. Ponizovskaya

*Laboratorio de Fisica de Sistemas Pequeños y Nanotecnología,*

*Consejo Superior de Investigaciones Cientificas,*

*Serrano 144, Madrid 28006, Spain*

M. Nieto-Vesperinas

*Instituto de Ciencia de Materiales de Madrid, Consejo*

*Superior de Investigaciones Científicas,*

*Campus de Cantoblanco Madrid 28049, Spain*

# Abstract

Recently it has been shown that due to losses the claims on observation of negative refraction using metamaterials are questionable. In this paper we discus an interesting case that is based in almost zero permeability at a given frequency: *zero permeability materials*. This represents a way out of the problem of left handed materials, and constitutes a material that acts a band pass filter with a very narrow band. The metamaterials claimed to present negative refraction are just showing this effect. However, absorption can be diminished in those metamaterials by using thicker wires. Materials based in quasi metallic magnetic structures are proposed with ferromagnetic resonance at microwave frequencies under applied magnetic fields. These materials, as well as the zero permitivity in thin films, open a new ground for a new set of optical devices.

Recently, it was shown that claims of observation of negative refraction[1] in developed "left-handed" metamaterials (LHM), initially proposed theoretically[2], are unfounded and misleading[3]. This is due to important losses that give rise to inhomogeneous waves whose propagation is dominated by damping within one wavelength $\lambda$. In addition, it was proven[4] that proposals of perfect lens slabs, even of hypothetical dispersiveless and absorptionless LHMs[5], are fundamentally wrong.

Specifically, a left-handed material has negative permittivity $\epsilon(\omega)$, permeability $\mu(\omega)$ and refractive index $n(\omega)$, where $\omega$ is the angular frequency. However, since the electromagnetic energy must be positive, the medium must necessarily be dispersive[2], which in turn implies absorption[6], namely, $Im[n(\omega)] > 0$ and the damping and losses dominate, so that one is in a territory of physics of metals. In fact, it is the condition $Im[n(\omega)] > 0$, for the wave moving towars infinity, what dominates the physical behavior of the wave transmitted into the medium, and thus rules over the sign that $Re[n(\omega)]$ might have[3]. Only if $Im[n]$ is small enough, in such a way that actual propagation of the wave exists in the medium, then the transmittivity through a slab of thickness $d$, shows the characteristic oscillating pattern versus frequency, with damping smaller than $1/e$. Uniquelly in this case it makes sense to talk of negative refraction when $Re[n] < 0$. However, this situation does not exist in any substance studied so far as LHM, whether metamaterial or not, because the strip wires used are 0.003cm-thick[7,8,9] and the looses are inversely proportional to the cross section of the wires.

The physical behavior of the permeability that we now address pertains to the region: $f < 10^{11}$–$10^{12} Hz$, ($\omega = 2\pi f$), where it can exhibit resonances. Notice that at frequencies above those values, $\mu(\omega)$ is unity for any material[6] or has deviations from unity that are negligible for the transmission problem at hand. Now, in the proximity of resonances $\mu$ may change sign and thus become zero at a certain frequency $\omega_c$. *Thus, we reach a situation of a medium with zero-like permeability ($z\mu M$).* Namely, for a small range of frequencies around $\omega_c$, one has: $Re[\mu(\omega_c)] \approx 0$ and $Im[\mu(\omega_c)]$ be small. This implies a region of frequencies where the refractive index is very small. Given the large slope of $Re[\mu]$ about this zero value, the wave beam will be highly collimated in frequency due to the abrupt exponential decay that it will suffer when its frequency slightly departs from $\omega_c$. Such a new material opens a new ground for optical experiments and devices.

On the other hand, at higher frecuencies a similar effect takes place for the dielectric permittivity. This would correspond to a *zero permittivity material ($Z\epsilon M$)*, already discussed in the context of photonic band gaps[7].

Let us consider the amplitude of the transmittivity $t$ of a slab of thickness $d$ and constitutive parameters $\epsilon$ and $\mu$, on which an s-polarized electromagnetic wave incides from vacuum. (The case of a p-polarized wave is treated in an analogous way). $t$ reads, (cf. Eq. (4) of Ref. 3):

$$t = \frac{\frac{4q_i q}{\mu} e^{-iq_i d}}{\left(q_i + \frac{q}{\mu}\right)^2 e^{-iqd} - \left(q_i - \frac{q}{\mu}\right)^2 e^{iqd}} \quad (1)$$

with $q_i = 2\pi \frac{\cos\theta_i}{\lambda}$ and $q = 2\pi n \frac{\cos\theta}{\lambda}$, $\lambda$ being the wavelength of the incident radiation. The transmittivity is $T = |t|^2$. We address normal incidence $\theta_i = \theta = 0$, like in the experiments of[1]. We shall denote frequencies $f$ in units of $10^{10} Hz$.

In the range of microwave frequencies for metals, or for structures with metallic elements, (let the last be either random or periodic lattices of e.g. grains or wires, immersed in dielectric media, or artificial crystals with mixed elements like those of Ref.1), the permittivity does not exhibit a resonant behavior, for the relevant frecuencies. We consider that the wires are metallic cylinders placed with separation $a = 0.5 cm$, and radius 0.0015 cm as in Refs. 1 and 8. There is a cutoff in $T$ at the frequency $f = c/2a = 3$ ($c$ being the speed of light). where we assume that the dielectric is vacuum. For the very narrow interval of frequencies which we will discuss (Fig. 1), the effective $\epsilon$ is practically constant and hence frequency independent. On the other hand, a different matter is the permeability $\mu(\omega)$ that has a resonant shape in those materials of interest for this discussion. i.e., we consider media that have $\mu$ of the following form:

$$\mu(\omega) = 1 - \frac{\omega_m^2 - \omega_0^2}{\omega^2 - \omega_0^2 + i\gamma\omega}(2) \qquad (1)$$

where $\omega_m$ and $\omega_0$ are the magnetic plasma and resonance angular frequencies, respectively, and $\gamma$ is the damping constant.

In Fig.1 we show the real and imaginary parts of $\mu(\omega)$, exhibiting a region of frequencies where $Re[\mu] < 0$. Then, one might argue that, since $Re[\epsilon] < 0$, and therefore $n = \sqrt{\epsilon\mu}$ has $Re[n] < 0$, that structure corresponds to LHM. However, both $Im[\epsilon]$ and $Im[\mu]$ are unavoidable large enough to make $Im[n]$ so great that it prevents wave propagation, in other words, inside the medium the electromagnetic wave is inhomogeneous, exponentially decaying i.e., the optical properties of the metamaterial are determined by the dominant $Im[n]$, and thus $Re[n]$ plays no role in practice, it is irrelevant whether it is positive negative.

There is, however, an interesting case, *when $Re[\mu] = 0$, i.e. when $\omega = \omega_c \approx \omega_m$*, this frecuency is of from the resonances frecuency $\omega_0$ where there is a very strong absorbing peak. Then, also $Im[\mu]$ is small enough for reasonable values of the damping constant $0.02 < \gamma < 0.2$, as obtained from the resistivity data of Cu and the section of the wires, (it should be noticed that these values of $\gamma$ are between 20 and 200 times larger than the value used in Ref. 1). Therefore in the previous conditions we have $\mu \approx 0$ and $n \approx 0$ and it can be proved that then the transmittivity $T$ has a very pronounced narrow peak (see Fig 1, and also Ref. 5). In fact, this is the only peak that $T$ has versus $\omega$. Accordingly, we call such a structure a *zero permeability material* (**Z$\mu$M**). Fig.1 shows this behavior for some characteristic parameters ($\epsilon$, $\omega_m/2\pi = 1.1$, $\omega_0/2\pi = 0.9$, $\gamma = 0.02$, and $d = 15cm$, as in Ref. 8). At this point we should say that, from finite-difference time-domain calculations, for the case at hand we found that the homogeneuos effective $\epsilon$ is aproximately equal to $-0.1 + 0.5i$[9]. Notice that its real part is practically zero, this is the case that agrees well with the experiments of[1] and[8] that showed a symmetric peak[8]. For example, Fig 1 shows that when $Re[\epsilon]$ is 0.5 or -0.5 there is a strong asimmetry in the peak of $t$, which is in contradiction with the experimental peak of[8]. When $d$ varies at the fixed frequency $\omega = \omega_c$ where $T$ has its maximum, then $T(\omega = \omega_c, d)$ decreases as $d$ increases, however, the interesting feature is that the width of the peak of $T$ around $\omega = \omega_c$ very drastically reduces with increasing $d$. Thus this transmittivity behaves as a narrow band pass filter. (Notice that had the numerator of Eq.(2) the form $\omega_m^2$, very similar things happen, but now at $\omega_c^2 = \omega_m^2 + \omega_0^2$).

In the case of a prism-shaped sample, like the one of Ref.1, the strong absorption in the medium gives rise at short distances from the sample, (i.e., not in the far zone), to a distribution of transmitted intensity which is concentrated in the thinner region of the prism. This well known effect ([10]) has been misinterpreted as negative refraction in[1]. However, the same phenomenon appears in any prism-shaped metallic sample. Specifically, we have performed experiments showing this effect in a prism-cut sample of Au, illuminated with visible light (750 − 380 nm)[11]; in this situation, $\mu = 1$.

We wish to remark that all reported experiments we are aware of, showed a narrow peak of intensity, with a value about $10^{-2}$ or $10^{-3}$ the incident intensity at frequency $\omega_c$. This peak very fastly decays when the frequency departs from $\omega_c$. This is exactly the behavior shown by Fig.1. However, this peak does not correspond neither to a left-handed nor to a right-handed medium. This is because of the two following reasons: first, due to the aforementioned dominant role of $Im[n]$. In fact, the wave in the material is quasistatic at the maximum $T(\omega \approx \omega_c)$, and, second, because the left-hand side of the peak is in the region $Re[\mu] < 0$, whereas the right-hand side of this peak is in the zone $Re[\mu] > 0$, (Fig. 1). So far, we have been discusing the experiments for the radius of the strip wires used experimentally of 0.003 cm thick[8].

As said about the losses are inversely proportional to the cross section of the wires. Our finite-diference time-domain simulations show that for these wires of 0.003cm the imaginary part of the permitivity is important and then there is no propagation in the LHM. However there may be a way to reduce the losses by increasing the wire radius by a factor of 10-20; i.e. for radii $\sim 0.05$ cm.

We should also stress that this behavior of those built metamaterials has nothing special, and certainly they do not represent a new state of matter as it has been claimed. For example, magnetic materials, whether nanostructured or not, at ferromagnetic resonance, have the same behavior as the former. Experiments in these last materials should show the same effects, but now at the frequency $\omega_c \approx \omega_m + \omega_0$, where $\omega_m = \gamma_M M$, and $\omega_0 = \gamma_M H$, where $\gamma_M$, $M$ and $H$ stand for the giromagnetic ratio, the magnetization of the sample, and the applied magnetic field, respectively.. The reason for this is that the permeability response of these materials is very similar to the one described by Eq.(2); the difference now being that $Im[\mu]$ has a sign different than the one in eq.(2).

Finally we should point out that by combining dielectric materials, metallic particles and thin wires, it is possible to obtain *zero-like permittivity materials*(**ZεM**) with $\mu = 1$. This has been recently reported in the context of photonic band gaps[7], where they also play an important role.

In conclusion, we have shown the existence of interesting properties of propagation of light in composite or metallic media, where the permeability has a resonant structure, at the frequency $\omega_c$ such that $Re[\mu(\omega_c)] = 0$, and thus the Z$\mu$M also has n$\approx$

0. This permits to have reasonable values of $T$ in a very small range of frequencies, thus providing and excellent band pass filter in the 1-20 Ghz region.

# 1. Acknowledgements

Work supported by the Spanish DGICYT.

# References

1. R.A. Shelby, D. R. Smith and S. Schultz, *Science* vol 292, 77 (2001).

2. V. G. Veselago, Sov. Phys. Usp. 10, 509 (1968).

3. N. Garcia and M. Nieto-Vesperinas, Optics Lett. 27, 885 (2002)

4. N. Garcia and M. Nieto-Vesperinas, Phys. Rev. Lett. 88, 207403 (2002).

5. J.B. Pendry, Phys. Rev. Lett. 85, 3966 (2000).

6. L. D. Landau and E. M. Lifshitz, *Electrodynamics of Continuous Media*, Pergamon Press, New York, 1960.

7. N. Garcia, E. V. Ponizovskaya, and John Q. Xiao, Appl. Phys. Lett. 80, 1120 (2002)

8. R. A. Shelby, D.R. Smith, S.C. Nemat-Nasser, and S. Schultz, Appl. Phys. Lett. 78, 489 (2001).

9. N. Garcia and E. V. Ponizowskaya are preparing the manuscript: "Determination of effective permittivity from finite-difference time-domain calculations".

10. M. Born and E. Wolf, *Principles of Optics*, 7th Edition, Cambridge University Press 1999.

11. M. Sanz et al., to be published.

## List of Figure Captions

Fig. 1: (a) $Re[\mu]$ and $Im[\mu]$ versus frequency. $Re[\mu] = 0$ at $f = f_m = 1.1$. At this value of $f$ $Re[\mu] \approx 0$ (see (b)) and $Im[n]$ is small (see (c)). Notice that the branch should always be taken corresponding to $Im[n] > 0$, as indicated by arrows. (d) shows the values of $T$ for $\epsilon = -0.1 + i0.5i$ (solid line); $\epsilon = -0.5 + i0.5i$ dashed line) and $\epsilon = 0.5 + i0.5i$ (dotted line). The calculated value of $\epsilon$ is: $\epsilon = -0.1 + i0.5i$[9] and is in excellent agreement with the experiment of[8]. The other two values of $\epsilon$ are excluded from the symmetric peak in the experiments[8]. Notice also that for $f < 1.1$ $Re[n] < 0$, and for $f > 1.1$ $Re[n] < 0$, at $f_c \approx 1.1$ $Re[n] = 0$.

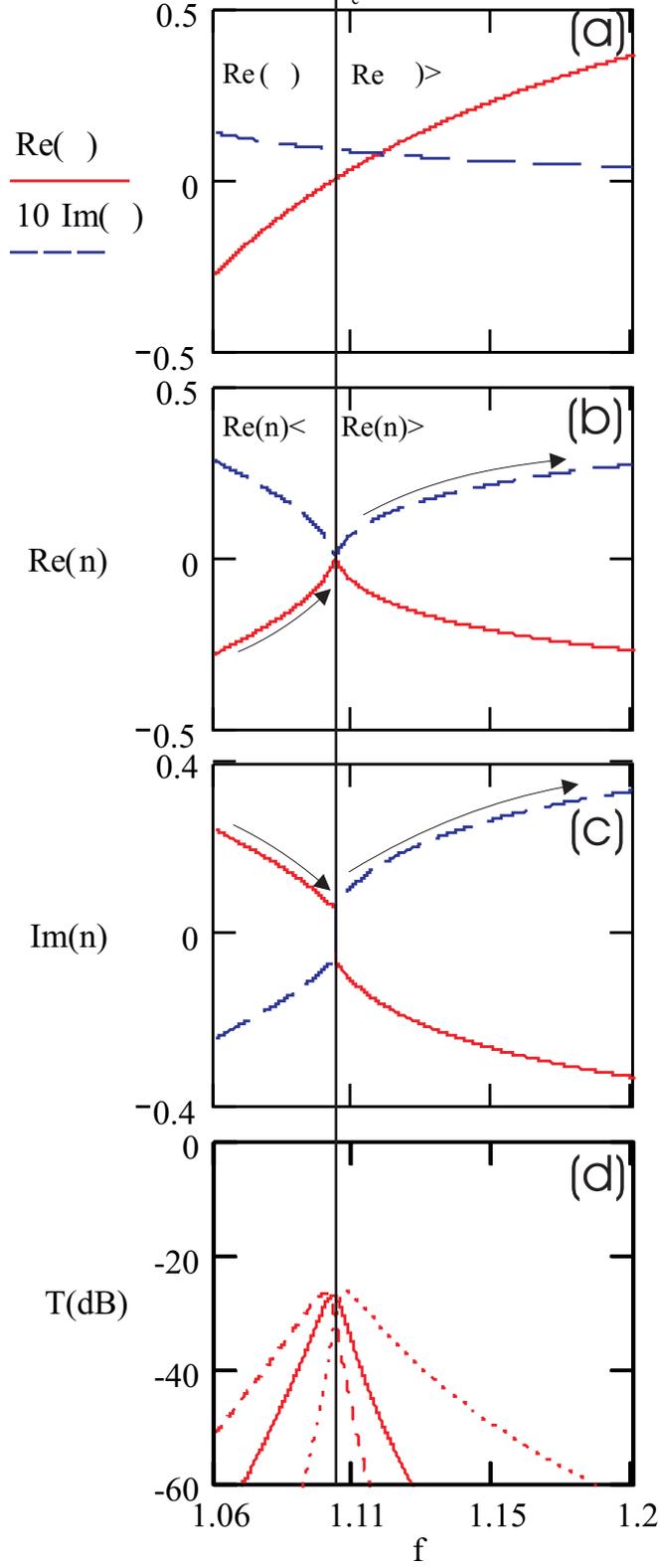

Fig. 1